# Quantum Imaging of Magnetic Phase Transitions and Spin Fluctuations in Intrinsic Magnetic Topological Nanoflakes

Nathan J. McLaughlin,[1] Chaowei Hu,[2] Mengqi Huang,[1] Shu Zhang,[2] Hanyi Lu,[1] Hailong Wang,[3] Yaroslav Tserkovnyak,[2] Ni Ni,[2] and Chunhui Rita Du[1,3]*

[1]Department of Physics, University of California, San Diego, La Jolla, California 92093
[2]Department of Physics and Astronomy, University of California, Los Angeles, California 90095
[3]Center for Memory and Recording Research, University of California, San Diego, La Jolla, California 92093

*Corresponding author: c1du@physics.ucsd.edu

Topological materials featuring exotic band structures, unconventional current flow patterns, and emergent organizing principles offer attractive platforms for the development of next-generation transformative quantum electronic technologies. The family of $MnBi_2Te_4$ $(Bi_2Te_3)_n$ materials is naturally relevant in this context due to their nontrivial band topology, tunable magnetism, and recently discovered extraordinary quantum transport behaviors. Despite numerous pioneering studies, to date, the local magnetic properties of $MnBi_2Te_4$ $(Bi_2Te_3)_n$ remain an open question, hindering a comprehensive understanding of their fundamental material properties. Exploiting nitrogen-vacancy (NV) centers in diamond, we report nanoscale quantum imaging of magnetic phase transitions and spin fluctuations in exfoliated $MnBi_2Te_4$ $(Bi_2Te_3)_n$ flakes, revealing the underlying spin transport physics and magnetic domains at the nanoscale. Our results highlight the unique advantage of NV centers in exploring the magnetic properties of emergent quantum materials, opening new opportunities for investigating the interplay between topology and magnetism.

**Introduction**

The interplay between magnetism and band topology promises to bring a plethora of exotic behaviors in emergent quantum materials, such as the quantum anomalous Hall effect (*1–6*), surface ferromagnetism (*7–10*), topological magnetoelectric effect (*11, 12*), and many others (*13–15*). Over the past few years, the family of $MnBi_2Te_4$ $(Bi_2Te_3)_n$ materials has emerged as an attractive platform to explore topologically dictated quantum spin and charge transport behaviors (*4, 16–30*). In contrast to magnetically doped topological systems, $MnBi_2Te_4$ $(Bi_2Te_3)_n$ harbors intrinsic magnetization carried by $Mn^{2+}$ ions in individual magnetic septuple layers and exhibits remarkable tunability in its macroscopic magnetic properties (*16, 17, 19, 31–34*). By controlling the odd-even layer-number determined (un)compensated magnetization, the interlayer coupling strength, and the external magnetic field strength, a set of distinct magnetic phases and electronic interactions can be realized in $MnBi_2Te_4$ $(Bi_2Te_3)_n$, leading to a variety of exotic quantum phenomena at relatively high temperatures (*4, 16, 18, 19, 22, 25, 27, 35*).

The discovery and understanding of these emergent material properties of two-dimensional (2D) $MnBi_2Te_4$ $(Bi_2Te_3)_n$ materials relies on concurrent developments in theory, material synthesis, and equally importantly, sensitive metrology tools capable of characterizing local magnetic properties at the nanoscale, which remains a formidable challenge in the current state of the art. The major difficulty results from the significantly reduced magnetic flux generated by the weak or nearly compensated magnetic moment of 2D $MnBi_2Te_4$ $(Bi_2Te_3)_n$ samples, which is beyond the detection limit of most conventional magnetometry techniques. While some other methods such as magneto-optic-Kerr-effect microscopy (*36*), reflective magnetic circular dichroism (*18*), and Raman spectroscopy (*30, 37*) could be employed to qualitatively diagnose the magnetic states of 2D $MnBi_2Te_4$ $(Bi_2Te_3)_n$ flakes, these techniques provide limited quantitative information about local magnetization and the field sensitivity of these tools is often compromised by interference effects and constrained by the optical diffraction limit. In this context, the local magnetic properties of 2D $MnBi_2Te_4$ $(Bi_2Te_3)_n$ samples has typically been inferred only indirectly from the electrical Hall or other measurement techniques in previous studies (*4, 18, 19, 21, 22, 31*).

To address this challenge, here, we introduce nitrogen-vacancy (NV) centers, optically active atomic defects in diamond that act as single-spin sensors (*38, 39*), to perform nanoscale quantum imaging of exfoliated $MnBi_2Te_4$ $(Bi_2Te_3)_n$ nanoflakes. By performing spatially resolved NV wide-field magnetometry measurements (*40–43*), we directly image the local magnetization of $MnBi_4Te_7$ nanoflakes and the characteristic multidomain evolution behavior in the first-order magnetic transition process. Exploiting the NV relaxometry method (*44–48*), we have observed spatially varied dynamic magnetic fluctuations in $MnBi_2Te_4$ $(Bi_2Te_3)_n$ nanoflakes, revealing underlying spin transport physics and the low-frequency spin dynamics hosted by domain walls in these magnetic materials. We expect that the presented NV quantum sensing and imaging platform can be readily extended to a large family of layered van der Waals crystals (*49–52*), providing ample opportunities for investigating the local spin dynamics and transport behaviors in a broad range of quantum materials (*48, 53–55*).

**Results**
**Experimental Details**

Figure 1A shows a schematic of our measurement platform, in which a [111] oriented diamond sample (*56*) containing shallowly implanted NV ensembles is used for the quantum sensing and imaging measurements. Note that one of the four possible NV orientations is along the out-of-plane direction (*z* axis), serving as an ideal local sensor to investigate the magnetic dynamics and phase transition of MnBi$_2$Te$_4$ (Bi$_2$Te$_3$)$_n$ with spontaneous perpendicular anisotropy. Microwave currents delivered by a freestanding Au wire are used to control the quantum spin state of NV centers, which can then be optically accessed via their spin-dependent photoluminescence. Our measurements start from the MnBi$_4$Te$_7$ compound, which exhibits an intermediate magnetic phase transition that occurs at low enough external fields to be fully accessible in our NV magnetometry system (*17*). An 83-nm-thick exfoliated MnBi$_4$Te$_7$ flake with approximate lateral dimensions of ~10 × 14 μm$^2$ is transferred onto the diamond surface, as shown in Fig. 1B (see Supplementary Materials Section 1 for details). Figure 1C illustrates the crystal and magnetic structure of MnBi$_4$Te$_7$, where the magnetic septuple layer of MnBi$_2$Te$_4$ and the nonmagnetic quintuple layer of Bi$_2$Te$_3$ stack alternately (*17*). In the magnetic ground state, moments carried by the Mn$^{2+}$ ions in individual septuple layers align parallel with each other along the out-of-plane direction due to the dominant intralayer ferromagnetic exchange interaction. Any two neighboring septuple layers are antiferromagnetically coupled, which can be further driven to a ferromagnetic configuration by applying a sufficiently strong external perpendicular magnetic field $B_{\text{ext}}$. Note that the intercalation by a nonmagnetic quintuple layer of Bi$_2$Te$_3$ increases the spacing between neighboring septuple layers, leading to a reduced interlayer exchange coupling strength and the relatively small first-order magnetic transition field (~1400 G) of MnBi$_4$Te$_7$ (*17*).

**Quantum imaging of the first-order magnetic phase transition of MnBi$_4$Te$_7$**

We first utilize NV spin sensors to image the field-driven antiferromagnetic-to-ferromagnetic phase transition of the exfoliated MnBi$_4$Te$_7$ flake. The d.c. NV wide-field magnetometry measurements exploit the Zeeman splitting effect of the NV $S = 1$ electron spins to measure the local magnetic field along the NV-orientation (*38*). The top panel of Fig. 1D shows the measurement protocol. We utilize 1-μs-long green laser pulses for NV initialization and readout and ~100-ns-long microwave $\pi$ pulses (*57*) to induce NV spin transitions. We sweep the frequency $f$ of the microwave $\pi$ pulses and measure the fluorescence across the field of view projected on a charge-coupled device (CCD) camera. When $f$ matches the NV electron spin resonance (ESR) frequencies, NV spins are excited to the $m_s = \pm 1$ states, which are more likely to relax through a non-radiative pathway back to the $m_s = 0$ ground state and emit reduced photoluminescence. The bottom panel of Fig. 1D shows an example of the measured NV ESR spectrum, which exhibits two pairs of split NV spin energy levels. The outer pair corresponds to the Zeeman splitting of NV centers with the out-of-plane (OOP) orientation while the inner one is degenerate, and corresponds to NV centers with axes along the other three possible directions. The magnitude of the local static magnetic field $B_{\text{tot}}$ can be extracted as follows: $B_{\text{tot}} = \pi \Delta f_{\text{OOP}}/\tilde{\gamma}$, where $\Delta f_{\text{OOP}}$ characterizes the Zeeman splitting of the OOP-oriented NVs, and $\tilde{\gamma}$ is the gyromagnetic ratio of the NV centers. Subtracting the contribution from the external magnetic field $B_{\text{ext}}$, the magnetic stray field $B_{\text{M}}$ generated from the MnBi$_4$Te$_7$ flake can be quantitatively measured. By performing spatially dependent optical detection of magnetic resonance (ODMR) measurements over the NV ensembles, we are able to obtain a 2D stray field map as illustrated by an example shown in Fig. 1E, which is measured at a temperature $T = 4.5$ K and $B_{\text{ext}} = 2296$ G. Through established reverse-propagation protocols (*41, 49, 50*) (see Supplementary Materials Section 2 for details), the corresponding magnetization ($4\pi M$) map of the exfoliated MnBi$_4$Te$_7$ flake can be quantitatively extracted, as shown in Fig. 1F.

To directly image the first-order magnetic phase transition, Figs. 2A-2F present the obtained 2D magnetization maps of the MnBi$_4$Te$_7$ flake measured with an increasing perpendicular magnetic field $B_{\text{ext}}$. In the low-field regime ($B_{\text{ext}} < 1000$ G), the exfoliated MnBi$_4$Te$_7$ flake exhibits a vanishingly small net magnetic moment, as shown in Fig. 2A and Fig. 2B, due to the antiferromagnetic coupling between the neighboring magnetic septuple layers. A finite magnetic moment shown in the top left region of the flake results from the field induced polarization. When $B_{ext}$ increases to 1220 G, the characteristic spin flip transition propagates from the top left area of the flake, leading to a "partial" ferromagnetic state with enhanced local magnetization. Such "partial" ferromagnetic phase features a co-existence of ferromagnetic and antiferromagnetic ordering of septuple layers along the out-of-plane direction due to the weak interlayer exchange coupling. Note that the remaining area of the MnBi$_4$Te$_7$ flake stays in the antiferromagnetic ground state with the formation of incipient magnetic domain walls at the locations where the energy barrier is lowest, as shown in Fig. 2C. The observed spin flip transition spreads out to the entire MnBi$_4$Te$_7$ flake through domain wall propagation with increasing perpendicular field (Fig. 2D and Fig. 2E) and the local magnetization further increases as the moments of the individual septuple layers become fully aligned. When $B_{ext} > 2000$ G, the magnetic phase transition of the MnBi$_4$Te$_7$ flake concludes with a saturated local magnetization, as shown in Fig. 2F.

Figure 2G summarizes the field-dependent evolution of the local magnetization at two different sites "A" and "B" of the exfoliated MnBi$_4$Te$_7$ flake, as marked in Fig. 2A. The magnetic moment at position "B" barely increases below the critical transition field, followed by a sudden jump of the net magnetic moment that saturates when $B_{ext}$ is larger than 1.8 kG. In contrast, the measured magnetization at the position "A" gradually increases with the perpendicular magnetic field. Around the spin flip transition point, the magnetic moment at the position "A" also features a step like jump followed by saturation in the high-field regime. The variation of the local magnetization of the MnBi$_4$Te$_7$ flake could result from the inhomogeneities (*31*), magnetic domains (*36*), and localized defects (*58*). The presented NV wide-field magnetometry results highlight a spatially nonuniform spin flip transition process of the 2D MnBi$_4$Te$_7$ flake at the nanoscale, exhibiting the characteristic magnetic domain nucleation and domain wall propagation behaviors. As a comparison, we also performed electrical Hall measurements to diagnose the field-dependent evolution of the magnetic phase of a patterned MnBi$_4$Te$_7$ flake device, as shown in Fig. 2H (see Supplementary Materials Section 1 for details). The measured Hall resistance manifests the typical bulk features (*17*) consisting of an initial plateau, followed by a spin flip transition and final saturation as the perpendicular magnetic field is increased.

**Quantum sensing of dynamic magnetic fluctuations of MnBi$_4$Te$_7$**

Known as a qubit-based magnetometer, NV centers possess excellent quantum coherence (*38*), providing opportunities for probing fluctuating magnetic fields that are challenging to access by conventional magnetometry methods (*44, 45, 47, 48*). Next, we apply NV wide-field magnetometry to directly measure the dynamic magnetic fluctuations in exfoliated MnBi$_2$Te$_4$ (Bi$_2$Te$_3$)$_n$ flakes, from which the intriguing physics underlying the spin transport and the low-frequency spin dynamics hosted by magnetic domain walls in this magnetic topological material family can be revealed. Figure 3A shows the dispersion of bulk MnBi$_4$Te$_7$ with an energy gap of ~51 GHz (see Supplementary Materials Section 3 for details). The magnon occupation follows the Bose-Einstein distribution. At thermal equilibrium, fluctuations of the transverse and longitudinal spin density of MnBi$_4$Te$_7$ generate fluctuating stray fields, as illustrated in Fig. 3B. Microscopically, the transverse spin fluctuations vanish below the magnon band minimum, and are related to the creation or annihilation of a single magnon (*48*, *59*). The longitudinal spin fluctuations are related to two-magnon processes (*46*, *54*, *55*), where a magnon with frequency $f$ is scattered to another one with frequency $f \pm f_{2m}$, emitting (absorbing) magnetic noise at frequency $f_{2m}$.

Experimentally, we used NV relaxometry technique to detect the fluctuating magnetic fields generated by the spin fluctuations of the $MnBi_4Te_7$ flake. The top panel of Fig. 3C shows the optical and microwave measurement sequence. A green laser pulse is first applied to initialize the NV spin to the $m_s = 0$ state. Spin fluctuations in the $MnBi_4Te_7$ sample couple to proximal NV centers through the dipole-dipole interaction. Fluctuating magnetic fields at the NV ESR frequencies induce NV spin transitions from the $m_s = 0$ to the $m_s = \pm 1$ states, leading to an enhancement of the NV relaxation rates $\Gamma$ (*44–48*). After a delay time $t$, we measure the occupation probabilities of the NV spin at the $m_s = 0$ and the $m_s = \pm 1$ states by applying a microwave $\pi$ pulse at the corresponding ESR frequencies and measuring the spin-dependent photoluminescence via the green-laser readout pulse. By measuring the integrated photoluminescence intensity as a function of the delay time $t$ and fitting the data with a three-level model (*48*, *59*), NV relaxation rates can be quantitatively obtained (see Supplementary Materials Section 4 for details). We highlight that in the low-field regime with a quasi-uniform magnetic domain picture, the NV ESR frequencies in our measurements are much lower than the minimum magnon energy of $MnBi_2Te_4$ $(Bi_2Te_3)_n$, thus, the measured NV relaxation is driven by the longitudinal spin fluctuations. In contrast, in the high-field regime where multidomain forms during the spin flip transition of $MnBi_2Te_4$ $(Bi_2Te_3)_n$, the magnetic noise generated by the low-frequency spin dynamics hosted by magnetic domain walls (*47*, *60*) also plays a role.

We first focus on NV relaxometry measurements in the low-field regime to detect the longitudinal spin fluctuations of the exfoliated $MnBi_4Te_7$ flake. Figures 3C-3E show the 2D maps of the measured NV relaxation rate at 8.5 K with characteristic ESR frequencies of 1.0 GHz, 1.2 GHz, and 2.7 GHz, respectively. The applied perpendicular magnetic field $B_{ext}$ is limited below 800 G to ensure the quasi-single magnetic domain feature of the sample. Notably, the measured NV spin relaxation rate $\Gamma$ is clearly enhanced in the diamond area that is covered with the $MnBi_4Te_7$ flake. The magnitude of $\Gamma$ decreases with increasing NV ESR frequency, suggesting that a larger energy difference invoked in the two-magnon process as discussed above will suppress longitudinal spin fluctuations, in qualitative agreement with our theoretical model (see Supplementary Materials Section 5 for details) (*54*, *55*). Figures 3F-3H show the 2D NV relaxation maps measured with the ESR frequency of 2.7 GHz at 4.5 K, 13 K, and 19 K, respectively. When $T = 4.5$ K, the magnetic fluctuations are largely suppressed due to the reduced thermal magnon energy, leading to negligible NV spin relaxation. Notably, we observed significantly enhanced longitudinal spin fluctuations at a temperature of 13 K, which is expected due to the increase in the magnetic susceptibility of $MnBi_4Te_7$ around the Néel temperature $T_N$ (*17*). When $T = 19$ K, longitudinal spin fluctuations remain active, which is attributed to the finite spin-spin correlation in the paramagnetic state (*61*). Note that the observed spatially varying magnetic fluctuations over the exfoliated $MnBi_4Te_7$ flake could be induced by inhomogeneities in magnetic susceptibility, spin diffusion constant, and exchange coupling strength as discussed below.

**Extraction of spin diffusion constant of $MnBi_2Te_4$ $(Bi_2Te_3)_n$ from NV relaxometry results**

The measured NV relaxation rate driven by longitudinal spin fluctuations at the NV ESR frequency $f_{ESR}$ is related by the fluctuation-dissipation theorem to the dynamic spin susceptibility $\chi''$ of the MnBi$_4$Te$_7$ sample and the NV transfer function in the high temperature limit as follows (54, 55):

$$\Gamma(f_{ESR}) = \frac{(\gamma\tilde{\gamma})^2 k_B T}{2\pi f_{ESR}} \int \mathcal{F}(\mathbf{k}, d)\chi''(\mathbf{k}, D)d\mathbf{k} + \Gamma^0 \qquad (1)$$

where the first term on the right represents the NV relaxation rate $\Gamma^M$ induced by spin dynamics of the MnBi$_2$Te$_4$ (Bi$_2$Te$_3$)$_n$ flake and $\Gamma^0$ is the intrinsic NV relaxation rate that is independent of the spin dynamics of MnBi$_2$Te$_4$ (Bi$_2$Te$_3$)$_n$. $\gamma$ and $\tilde{\gamma}$ are the gyromagnetic ratio of the magnetic sample and NV centers, $k_B$ is the Boltzmann constant, $T$ is temperature, $\mathbf{k}$ is the magnon wavevector, $\mathcal{F}(\mathbf{k}, d)$ is the transfer function (59) describing the magnetic fields generated at the NV site and $d$ is the NV-to-sample distance. The dynamic spin susceptibility $\chi''$ reflects the underlying spin transport properties of a magnetic sample, which can be expressed as a function of the intrinsic spin diffusion constant $D$ and the static longitudinal magnetic susceptibility per layer $\chi_0$ via the conventional diffusion equation (55, 62) (See Supplementary Materials Section 5 for details). Therefore, by measuring the NV relaxation rate as a function of ESR frequencies, we are able to extract the key material parameters $D$ and $\chi_0$ of the MnBi$_4$Te$_7$ flake. Figure 4A plots a three-dimensional map of the spatially averaged NV relaxation rates $\Gamma^M$ of the flake as a function of ESR frequency and temperature. The variation of the NV relaxation rate agrees well with the theoretical model of Eq. (1), from which the intrinsic spin diffusion constant $D$ and the static longitudinal magnetic susceptibility $\chi_0$ of MnBi$_4$Te$_7$ at individual temperatures can be extracted, as presented in Figs. 4B and 4C. $D$ is obtained to be $(6.1 \pm 0.8) \times 10^{-6}$ m/s$^2$ at 4.5 K, in qualitative agreement with the theoretical estimation of $D = v^2\tau/2$ (63) by taking the magnon velocity $v \sim 1$ km/s (17) and the momentum scattering time $\tau \sim 10$ ps. Below the Néel temperature (~13 K) of MnBi$_4$Te$_7$, $D$ monotonically decreases as the temperature increases due to a reduction of $\tau$ caused by enhanced Umklapp scattering (64). When $T > 14$ K, $D$ decays to an approximately constant value and the residual spin transport capability results from atomic scale thermal spin fluctuations. The obtained static magnetic susceptibility $\chi_0$ generally agrees with the previous magnetometry measurement of the bulk MnBi$_4$Te$_7$ samples (17). As shown in Fig. 4C, $\chi_0$ reaches a peak value of $(9.9 \pm 0.6) \times 10^{-3}$ nm around the Néel temperature of MnBi$_4$Te$_7$, which is consistent with the antiferromagnetic phase transition (17). Figures 4D and 4E show the obtained 2D maps of spin diffusion constant $D$ and static magnetic susceptibility $\chi_0$ of the exfoliated MnBi$_4$Te$_7$ flake, respectively, measured at 8.5 K. Note that the enhanced $\chi_0$ in the top left region of the flake is consistent with the observed larger field-induced magnetic moment, as shown in Figs. 2A and 2B.

To reveal the underlying spin transport mechanism in the magnetic topological material MnBi$_2$Te$_4$ (Bi$_2$Te$_3$)$_n$ family, we extend the demonstrated NV wide-field relaxometry platform to MnBi$_2$Te$_4$ and MnBi$_8$Te$_{13}$. Figures 5A-5C show schematics of the material structure of MnBi$_2$Te$_4$, MnBi$_4$Te$_7$, and MnBi$_8$Te$_{13}$, respectively. In MnBi$_2$Te$_4$, magnetic septuple layers directly stack on each other without the insertion of the nonmagnetic quintuple layer (Fig. 5A). The higher-order magnetic topological compounds MnBi$_4$Te$_7$ and MnBi$_8$Te$_{13}$ feature an increasing number of the intercalating quintuple layers (16), as illustrated in Figs. 5B and 5C. By changing the spacing between neighboring magnetic septuple layers, the interlayer coupling strength of MnBi$_2$Te$_4$ (Bi$_2$Te$_3$)$_n$ systematically varies, while the intralayer coupling strength remains largely the same (*65*). Figures 5D-5F present the 2D maps of spin diffusion constant of the exfoliated MnBi$_2$Te$_4$, MnBi$_4$Te$_7$, and MnBi$_8$Te$_{13}$ flakes measured at 8.5 K, respectively. The thicknesses of the exfoliated flakes have been characterized by atomic force microscopy (See Supplementary Materials Section 1 for details). Despite the difference in the shape of the flakes and defect-induced inhomogeneities, the spatially averaged spin diffusion constants of the three MnBi$_2$Te$_4$ (Bi$_2$Te$_3$)$_n$ samples show notable consistency, which are $(5.7 \pm 0.6) \times 10^{-6}$ m/s$^2$, $(5.4 \pm 0.7) \times 10^{-6}$ m/s$^2$, and $(5.4 \pm 0.4) \times 10^{-6}$ m/s$^2$, respectively, suggesting a common mechanism underlying the intrinsic spin transport in this magnetic material family. Microscopically, the spin diffusion process describes the random Brownian motion of a thermally populated magnon gas with a velocity $v$ governed by the exchange energy. The intrinsic spin diffusion constant is determined by: $D = v^2\tau/2$ (*63*), where $v \sim (J_s a)/\hbar$ (*66*) ($J_s$ is the exchange coupling strength, $a$ is the lattice constant, and $\hbar$ is the reduced Planck constant) and $\tau$ is the momentum relaxation time. The minimal variation of the obtained $D$ of MnBi$_2$Te$_4$, MnBi$_4$Te$_7$, and MnBi$_8$Te$_{13}$ flakes indicates that spin diffusion in MnBi$_2$Te$_4$ (Bi$_2$Te$_3$)$_n$ family is mainly driven by the intralayer exchange interaction while the role of the interlayer coupling is secondary. Intuitively, we expect that spin transport in MnBi$_2$Te$_4$ (Bi$_2$Te$_3$)$_n$ is largely confined in individual magnetic septuple layers and the effect of interlayer spin diffusion is negligible, as illustrated in Figs. 5A-5C.

**Quantum imaging of magnetic domain walls formed in an MnBi$_4$Te$_7$ flake**

In magnetic topological materials, domain walls serve as an intrinsic boundary between two topological states with opposite chiralities (*15, 67*). Because the Chern number must change discontinuously across neighboring magnetic domains, chiral edge states are expected to reside at the domain wall sites, offering an attractive platform to realize and engineer exotic quantum spin and charge transport behaviors (*15, 67*). In the above sections, we have demonstrated NV wide-field imaging of the longitudinal spin fluctuations generated by a quasi-uniform magnetic domain in MnBi$_4$Te$_7$. Next, we move to the high-field regime, where the spin flip transition occurs, to image spin fluctuations generated by magnetic domain walls in an MnBi$_4$Te$_7$ flake. Figure 6A illustrates the evolution of magnetic phases of MnBi$_4$Te$_7$ as a function of increasing perpendicular magnetic field $B_{ext}$. In the antiferromagnetic and ferromagnetic states, the magnetization of each septuple layer is uniformly aligned and the bulk spin excitation energy of MnBi$_4$Te$_7$ is larger than the NV ESR frequencies in the accessible magnetic field range due to the large magnetic anisotropy. In this case, the NV-magnon coupling is mediated by the longitudinal spin fluctuations as discussed above. During the spin flip transition process, spins in individual septuple layer may flip mostly independently due to the weak interlayer coupling. Nucleation of domain walls occurs at locations within of an exfoliated MnBi$_4$Te$_7$ flake where the energy barrier is the lowest, leading to the formation of a partially magnetically flipped septuple layer, as illustrated in the middle panel of Fig. 6A. The low-frequency spin dynamics hosted by magnetic domain walls will generate fluctuating magnetic stray fields (*47, 60*), driving NV relaxation at the corresponding ESR energy.

Figures 6B-6D show the NV wide-field imaging of static magnetization of the MnBi$_4$Te$_7$ flake measured at 1311 G, 1442 G, and 2340 G, respectively. A perpendicular magnetic field $B_{ext}$ is applied to induce formation of magnetic domains during the spin flip transition (Figs. 6B and 6C). Figures 6E-6G show 2D maps of the NV relaxation rate measured at the corresponding magnetic fields. Note that the background of the NV relaxation rate induced by the longitudinal

spin noise generated from the bulk spin excitation has been subtracted to highlight the contribution $\Gamma^D$ from the spin fluctuations in magnetic domain walls. (See Supplementary Materials Section 6 for details). When $B_{ext}$ = 1311 G, the top-left corner of the MnBi$_4$Te$_7$ flake first becomes polarized while the rest of the sample remains partially magnetically flipped. This is accompanied by the formation of lateral magnetic domain walls that generate significantly enhanced magnetic noise. Note that the flipped area of the MnBi$_4$Te$_7$ flake expands when increasing $B_{ext}$ to 1422 G, leading to the suppression of the NV relaxation rate, as shown in Fig. 6F. When $B_{ext}$ = 2340 G, MnBi$_4$Te$_7$ enters the fully saturated ferromagnetic state, as shown in Fig. 6D, generating a vanishingly small spin noise $\Gamma^D$ due to the absence of magnetic domain walls. In a qualitative picture, the measured NV relaxation rate increases with the domain wall density, which is further related to the gradient of the magnetization (See Supplementary Materials Section 7 for details).

Direct imaging of magnetic domain walls in intrinsic magnetic topological materials could provide an attractive platform to investigate the interplay between the local magnetization and the macroscopic quantum transport behaviors, e.g. the high-temperature quantum anomalous Hall effect (*4–6*, *25*, *68*) and layer Hall effect (*19*) recently discovered in MnBi$_2$Te$_4$ (Bi$_2$Te$_3$)$_n$. For the presented NV wide-field imaging technique, the spatial resolution lies in the range of ~500 nanometers, which is restricted by the optical diffraction limit. By employing scanning NV microscopy methods (*47*, *49*, *51*, *69*), we expect that the spatial resolution could ultimately reach the tens of nanometers regime, offering new opportunities to uncover detailed microscopic features of magnetic patterns in a broad range of quantum materials.

**Discussion**

In summary, we have demonstrated NV centers as a nanoscale local probe to investigate magnetic phase transitions and spin fluctuations of 2D magnetic topological materials. The spatially resolved NV magnetometry measurements reveal the characteristic domain wall nucleation and propagation features during the spin flip transition of MnBi$_4$Te$_7$ flakes. By employing NV relaxometry technique, we access the longitudinal spin fluctuations in the family of MnBi$_2$Te$_4$ (Bi$_2$Te$_3$)$_n$ materials in the low-field regime, highlighting the peculiar intralayer dominated spin diffusive behavior confined in the individual magnetic septuple layers. During the spin flip transition process, we observe a dramatic increase of magnetic noise arising from magnetic domain walls in an MnBi$_4$Te$_7$ flake, which is attributable to the characteristic low-frequency spin fluctuations hosted by the spatially evolving magnetic textures. Our results illustrate the unique capabilities of NV centers in probing the nanoscale spin transport and dynamics of MnBi$_2$Te$_4$ (Bi$_2$Te$_3$)$_n$, opening new possibilities for investigating the interplay between magnetism and topology in a broad range of low-dimensional materials. The dynamic coupling between topological materials and NV spin qubits may also point to the possibility of developing NV-based hybrid architectures for the next-generation quantum information sciences and technologies (*60*).

**Acknowledgements**. Authors would like to thank Eric E. Fullerton for providing the Physical Property Measurement System for the electrical transport measurements and Yuxuan Xiao and Eric Lee-Wong for assistance in sample preparation and magnetometry characterization. Authors thank Yu-Hang Li, Ran Cheng, Yi-Zhuang You, Suyang Xu, Sheng-Chin Ho, and Anyuan Gao for insightful discussions. N. J. M., H. W., H. L., and C. H. R. D. were supported by the Air Force Office of Scientific Research under award FA9550-20-1-0319 and its Young Investigator Program under award FA9550-21-1-0125. M. H. and C. H. R. D. acknowledged the support from U. S. National Science Foundation (NSF) under award ECCS-2029558 and DMR-2046227. C. H. and N. N. were supported by the U.S. Department of Energy (DOE), Office of Science, Office of Basic Energy Sciences under Award No. DE-SC0021117 for the bulk single crystal growth and characterization. S. Z. and Y. T. were supported by NSF under Grant No. DMR-1742928.

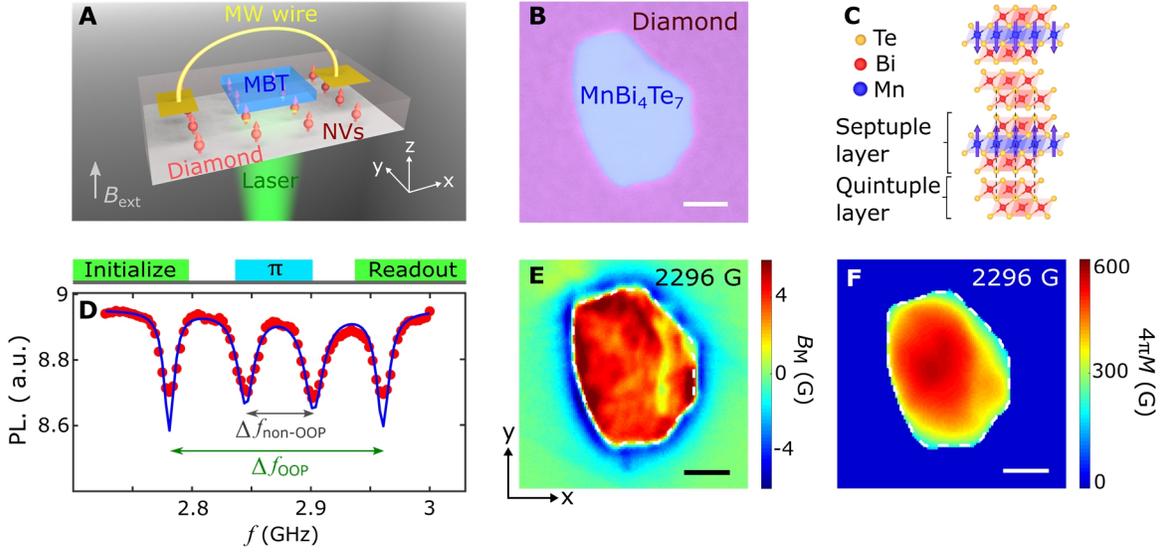

**Figure 1. Quantum imaging of static magnetic stray fields generated by an exfoliated MnBi$_4$Te$_7$ flake.** (**A**) Schematic of an exfoliated MnBi$_4$Te$_7$ flake transferred onto a [111] oriented diamond membrane for nitrogen-vacancy (NV) wide-field magnetometry measurements. (**B**) Optical image of a prepared NV-MnBi$_4$Te$_7$ device. (**C**) Schematic of the crystal and magnetic structures of MnBi$_4$Te$_7$ consisting of alternately stacked septuple and quintuple layers. (**D**) Optically detected magnetic resonance spectrum of NV centers contained in [111] oriented diamond with an external field $B_{ext}$ applied along the out-of-plane (OOP) direction. The experimental spectrum is comprised of two pairs of resonance peaks corresponding to the Zeeman splitting of NV spins with OOP orientation and NV spins that are aligned along other directions (non-OOP). (**E**)-(**F**) Two-dimensional images of static stray field $B_M$ (**E**) and reconstructed magnetization $4\pi M$ (**F**) of the exfoliated MnBi$_4$Te$_7$ flake measured at 4.5 K with $B_{ext}$ = 2296 G. The white dashed lines mark the boundary of the exfoliated MnBi$_4$Te$_7$ flake and the scale bar is 4 μm in Figs. (**B**), (**E**), and (**F**).

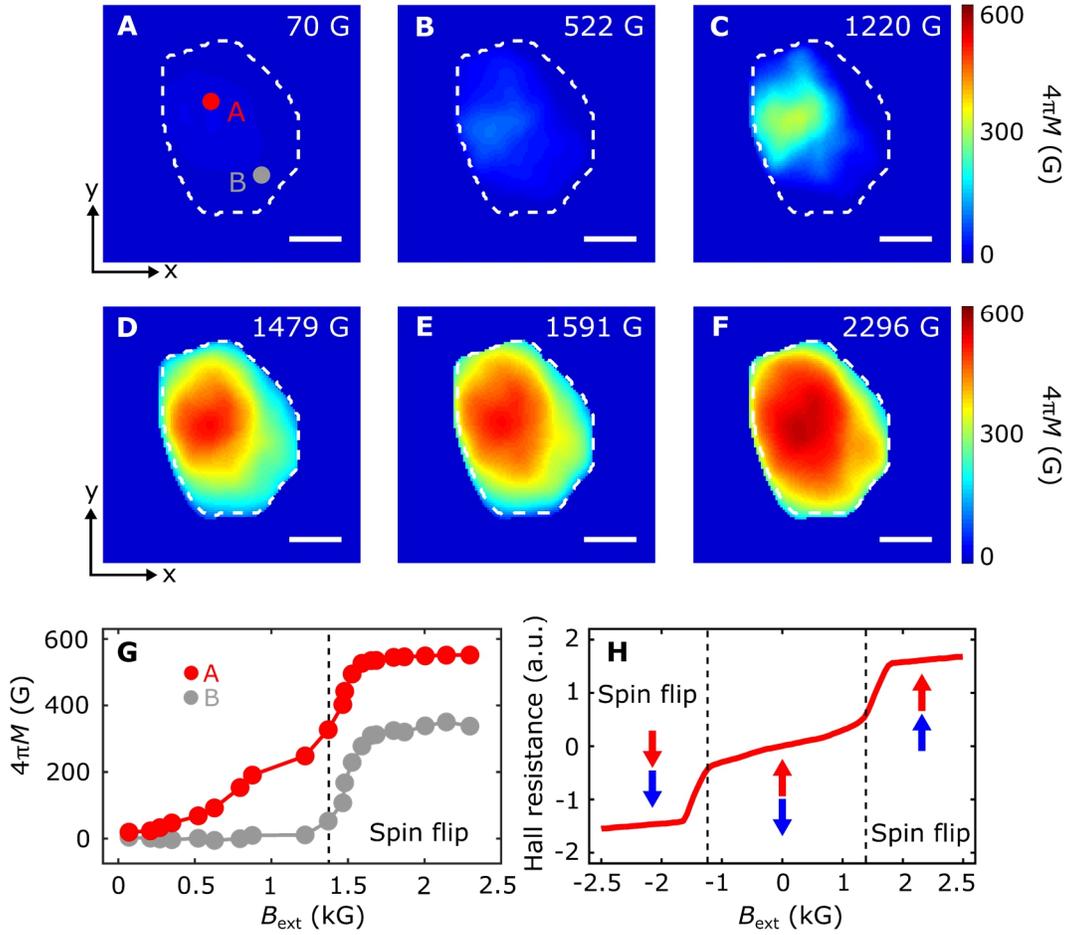

**Figure 2. Quantum imaging of the first order magnetic phase transition of an MnBi$_4$Te$_7$ flake.** (**A**)-(**F**) Magnetization ($4\pi M$) maps of an MnBi$_4$Te$_7$ flake measured at 4.5 K and $B_{ext}$ = 70 G (**A**), 522 G (**B**), 1220 G (**C**), 1479 G (**D**), 1591 G (**E**), and 2296 G (**F**), respectively. The white dashed lines mark the boundary of the exfoliated flake, and the scale bar is 4 μm. (**G**) Local magnetization measured at two different sites "A" (red point) and "B" (grey point) of the MnBi$_4$Te$_7$ flake as a function of the perpendicular magnetic field $B_{ext}$. The specific locations of the two points "A" and "B" are marked in Fig. 2(**A**). (**H**) Hall resistance of a prepared MnBi$_4$Te$_7$ device measured as a function of $B_{ext}$. The dashed lines indicate the critical field of the spin flip transition of MnBi$_4$Te$_7$ and the red (blue) arrows illustrate the magnetic configurations of two neighboring septuple layers in MnBi$_4$Te$_7$ in different field regimes.

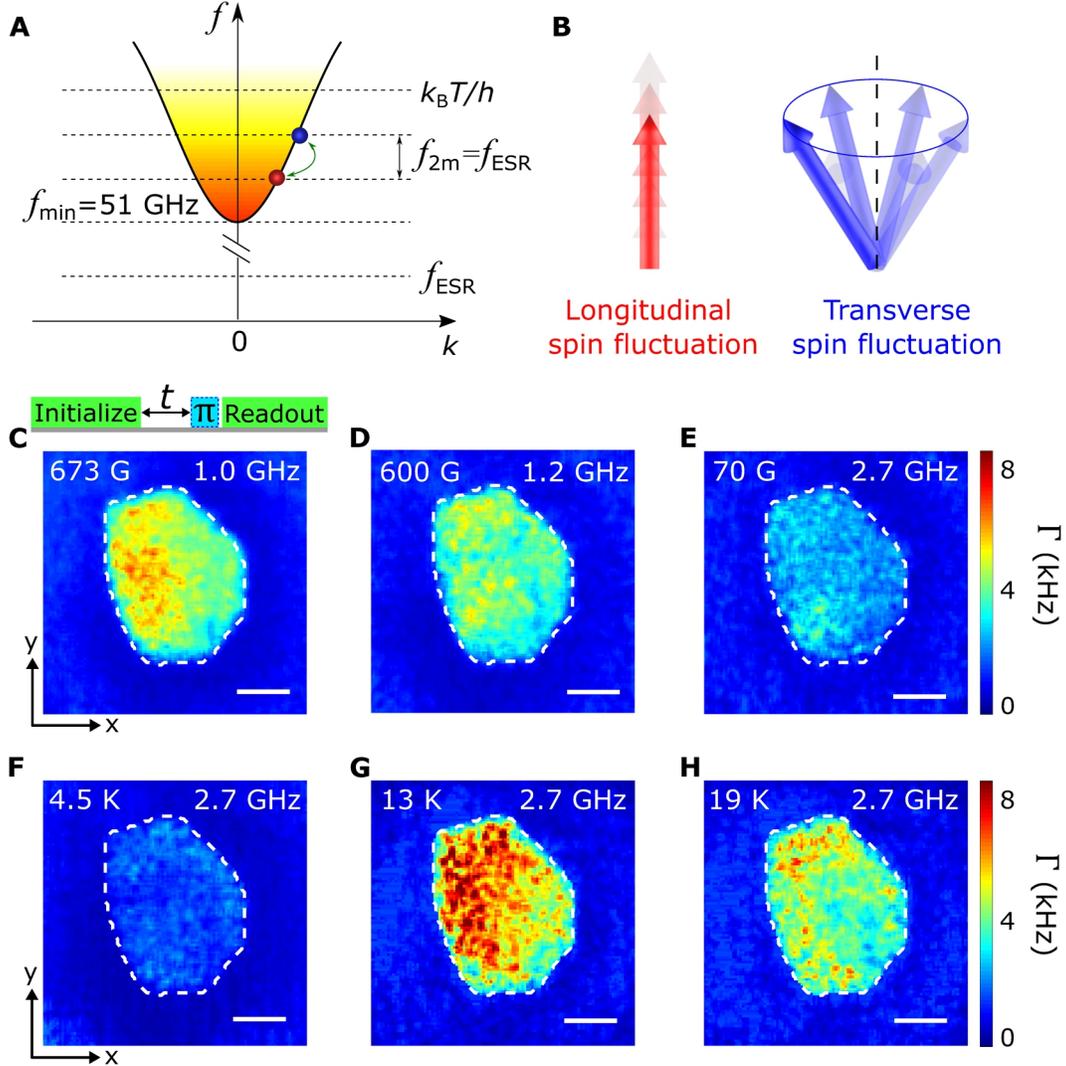

**Figure 3. Quantum sensing of longitudinal spin fluctuations of an MnBi$_4$Te$_7$ flake.** (**A**) Sketch of the magnon dispersion of bulk MnBi$_4$Te$_7$. The magnon occupation follows the Bose-Einstein distribution $1/[\exp\left(\frac{hf}{k_\mathrm{B}T}\right) - 1]$, as indicated by the fading colors. The minimal magnon frequency $f_\mathrm{min}$ is ~51 GHz, much larger than the NV electron spin resonance frequencies $f_\mathrm{ESR}$ in our measured field range. The longitudinal spin fluctuations with a characteristic frequency of $f_\mathrm{2m}$ are associated with two-magnon scattering processes, and couple with an NV center through dipolar interactions. (**B**) Schematic of the dynamic fluctuations of the longitudinal and transverse spin density. (**C**)-(**E**) NV spin relaxation maps measured at $f_\mathrm{ESR}$ of 1.0 GHz (**C**), 1.2 GHz (**D**), and 2.7 GHz (**E**) at a temperature of 8.5 K. Top panel of (**C**): optical and microwave sequence of NV relaxometry measurements. (**F**)-(**H**) NV spin relaxation maps measured at 4.5 K (**F**), 13 K (**G**), and 19 K (**H**) with $f_\mathrm{ESR}$ = 2.7 GHz. The white dashed lines mark the boundary of the exfoliated flake, and scale bar is 4 μm in Figs. (**C**)-(**H**).

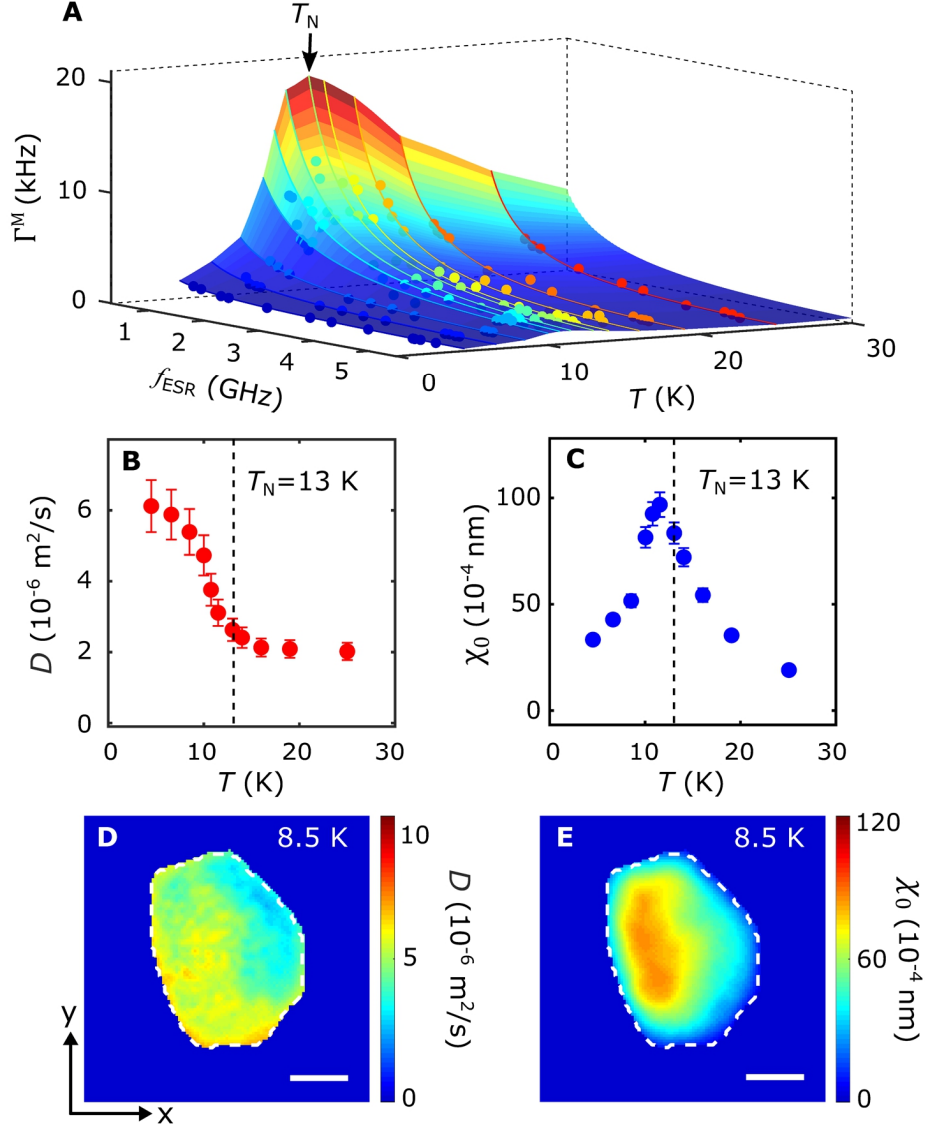

**Figure 4. Extraction of spin diffusion constant and magnetic susceptibility of an MnBi$_4$Te$_7$ flake.** (**A**) NV relaxation rates $\Gamma^M$ measured as a function of temperature and the NV electron spin resonance frequency $f_{ESR}$, from which the spin diffusion constant and magnetic susceptibility per layer can be extracted. The solid lines are fitting curves to the theoretical calculation. (**B**) Intrinsic spin diffusion constant $D$ measured as a function of temperature between 4.5 K and 25 K. (**C**) Longitudinal static magnetic susceptibility $\chi_0$ measured as a function of temperature between 4.5 K and 25 K. (**D**)-(**E**) Obtained 2D images of the spin diffusion constant and the longitudinal static magnetic susceptibility of the MnBi$_4$Te$_7$ flake at 8.5 K. The white dashed lines mark the boundary of the exfoliated flake, and the scale bar is 4 μm.

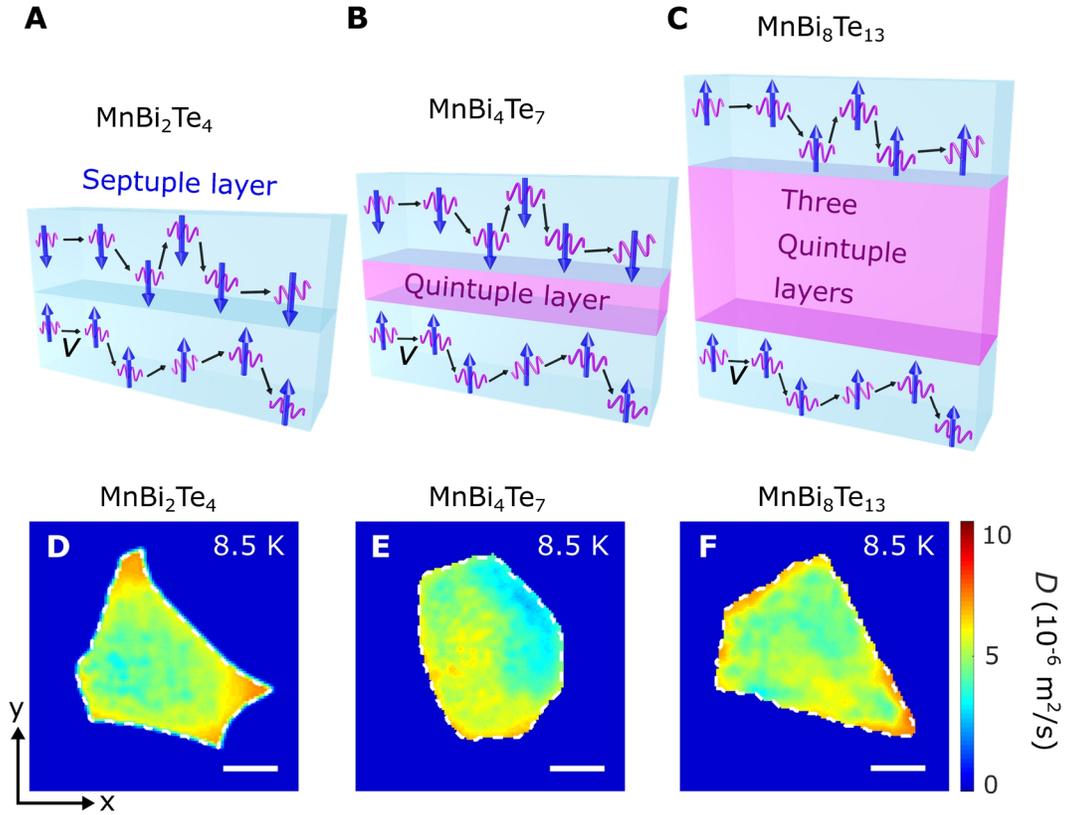

**Figure 5. Intralayer exchange coupling driven spin diffusion in the family of MnBi$_2$Te$_4$ (Bi$_2$Te$_3$)$_n$ materials.** (A)-(C) Schematics of the material structure of MnBi$_2$Te$_4$ (A), MnBi$_4$Te$_7$ (B), and MnBi$_8$Te$_{13}$ (C) consisting of alternately stacked nonmagnetic quintuple layers and magnetic septuple layers. Spin diffusive transport is driven by spin-conserved scattering of a thermally populated magnon gas with a characteristic velocity $v$ that is dominated by the intralayer exchange interaction. Blue arrows represent the spin carried by magnons (denoted by purple waves) in magnetic septuple layers. The spatially random trajectory of the spin (denoted by the black arrows) represents the exchange-dominated spin scattering process. (D)-(F) 2D maps of the obtained spin diffusion constant of exfoliated MnBi$_2$Te$_4$ (D), MnBi$_4$Te$_7$ (E), and MnBi$_8$Te$_{13}$ (F) flakes measured at 8.5 K. The white dashed lines mark the boundary of the exfoliated flakes, and the scale bar is 4 μm.

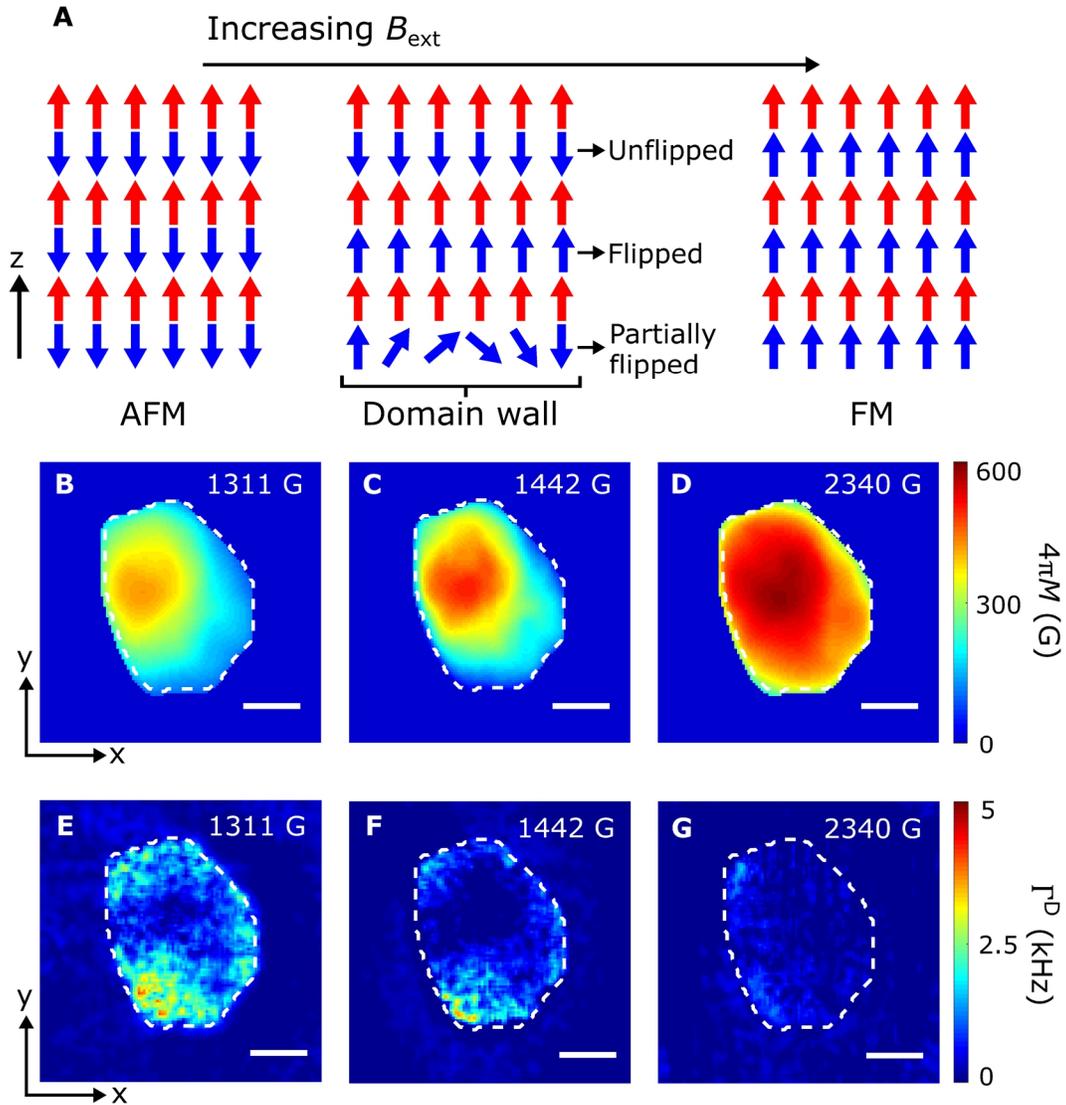

**Figure 6. NV wide-field imaging of spin fluctuations generated from lateral magnetic domain walls formed in an MnBi$_4$Te$_7$ flake.** (**A**) Schematic of the evolution of the magnetic states of MnBi$_4$Te$_7$ as a function of increasing perpendicular magnetic field $B_{ext}$. In the ferromagnetic and antiferromagnetic phases, the magnetic moments of individual magnetic septuple layers are fully aligned parallel or antiparallel without the formation of magnetic domains. During the spin flip transition process, spins in individual septuple layers could flip mostly independently. Magnetic domain walls nucleate at locations where the energy barrier is lowest, leading to magnetic gradient in the lateral direction of a partially flipped septuple layer. (**B**)-(**D**) Reconstructed magnetization maps of the MnBi$_4$Te$_7$ flake measured at $B_{ext}$ of 1311 G (**B**), 1442 G (**C**), and 2340 G (**D**). (**E**)-(**G**) Magnetic domain wall-induced NV spin relaxation maps measured at the corresponding magnetic fields. The longitudinal spin noise generated from the bulk spin excitations has been subtracted to highlight the contribution $\Gamma^D$ from the magnetic domain walls. The white dashed lines mark the boundary of the exfoliated flake, and the scale bar is 4 μm in Figs. (**B**)-(**G**).